\documentclass[aip,jcp,numerical,reprint]{revtex4-1}
\usepackage{graphicx}  
\usepackage{threeparttable} 
\usepackage{siunitx} 
\usepackage{bm,amssymb,amsmath} 
\usepackage[bookmarks]{hyperref} 
\usepackage[svgnames]{xcolor} 
\definecolor{Blue}{RGB}{0 0 128}
\hypersetup{ 
	colorlinks= true,
    allcolors = Blue
}
\usepackage{tikz}
\usetikzlibrary{calc} 
\usetikzlibrary{arrows,decorations.markings} 
\usepackage{color,soul} 

\usepackage{comment}
\begin{document}

\title{Interpolated wave functions for nonadiabatic simulations with the fixed-node quantum Monte Carlo method}
\author{Norm Tubman}
\email[electronic address: ]{ntubman@berkeley.edu}
\affiliation{University of California Berkley, Department of Chemistry, Hildebrand Hall, Berkeley CA, 94720}
\author{Yubo Yang}
\affiliation{University of Illinois Urbana-Champaign, Department of Physics, 1101 W Green Street, Urbana, IL 61801}
\author{Sharon Hammes-Schiffer}
\affiliation{University of Illinois Urbana-Champaign, Department of Chemistry, Urbana, IL 61801}
\author{David Ceperley}
\affiliation{University of Illinois Urbana-Champaign, Department of Physics, 1101 W Green Street, Urbana, IL 61801}

\begin{abstract}
Simulating nonadiabatic effects with many-body wave function approaches is an open field with many challenges.  
Recent interest has been driven by new algorithmic developments and improved theoretical understanding of properties unique to electron-ion wave functions.  Fixed-node diffusion Monte Caro is one technique that has shown promising results for simulating electron-ion systems.
In particular, we focus on the CH molecule for which previous results suggested a relatively significant contribution to the energy from nonadiabatic effects.  We propose a new wave function ansatz for diatomic systems which involves interpolating the determinant coefficients calculated from configuration interaction methods.    We find this to be an improvement beyond previous wave function forms that have been considered.    The calculated nonadiabatic contribution to the energy in the CH molecule is reduced compared to our previous results, but still remains the largest among the molecules under consideration.

\end{abstract}
\maketitle

\section{Introduction}
The Born-Oppenheimer approximation is widely used in the simulation of chemical and condensed matter systems~\cite{Tubman_ECG,Martinez_Review,Cederbaum_Review}. However, the breakdown of the Born-Oppenheimer approximation can lead to interesting new physics and in some cases even giant effects~\cite{saitta2008,giustino2016}. The full impact of using the Born-Oppenheimer approximation is still not widely understood due to the lack of theoretical methods that can go beyond the Born-Oppenheimer approximation accurately, although there has been significant progress recently made in this direction\cite{Tubman_ECG,Yang2015,Sharon_NEO-HF,Sharon_XCNEO-HF1,Sharon_XCNEO-HF2,Sharon_XCNEO-HF,Kurt_XCNEO-HF,Kurt_XCNEO-HF1,Sharon_NEO-DFT,Sharon_NEO-DFT2,Sharon_NEO-DFT3,Gross_NEO-DFT,Gross_NEO-DFT1,Ilkka_Path,Ilkka_Path1,Ilkka_Path2,Ceperley_1987}.  While nonadiabatic effects are ignored in many applications, there are several important places where highly accurate simulations that go beyond the Born-Oppenheimer approximation are imperative.  For example, the identification of molecules in the diffuse interstellar bands (DIB) is one such case where highly accurate energies without the Born-Oppenheimer approximation is needed for both ground and excited states~\cite{snow2006}. Recent experiments have been able to identify several peaks that correspond to ionized C$_{60}$ in the DIB\cite{campbell2015}.  However, many molecules remain unidentified and as such there are still many open questions as to the physical processes that occur in the interstellar medium.   Theoretical approaches have not been widely used to directly identify absorption lines in the spectrum due to a lack of accuracy in current simulations.  

While identifying absorption peaks in the DIB is beyond what we can simulate currently, we have started developing quantum Monte Carlo techniques to make progress in this direction\cite{Tubman_ECG,Yang2015}.   Our current focus has been to benchmark molecular systems in which the interactions between the electrons and ions are not approximated.   Fixed-node diffusion Monte Carlo (FN-DMC) is a method in which results are only biased by what is called the fixed-node approximation.  
The fixed-node approximation has been tested extensively and can be used to produce some of the best results for clamped-ion simulations~\cite{grossman1,Tubman_Release,Tubman_ACS}.  For FN-DMC simulations that go beyond the Born-Oppenheimer approximation, recent benchmarks demonstrated some of the most accurate energies ever calculated for a series of atomic and diatomic systems~\cite{Yang2015}.

There is, however, much development still needed to improve the accuracy and scalability of these simulations even further.  The nonadiabatic effects in the atomic and and diatomic systems, as calculated from FN-DMC simulations,  were generally smaller than 0.1~mHa. There were some exceptions, and in particular the CH molecule had an unexpectedly large contribution from nonadiabatic effects. While there could be some nonadiabatic effects in the CH molecule, it is difficult to determine exactly how much of the previous estimate is due to the fixed-node error.   To address this question further we develop a wave function form that is improved over those used in our previous simulations of CH.

\section{Fixed-Node Diffusion Monte Carlo (FN-DMC)}
Diffusion Monte Carlo~\cite{Anderson_DMC,lester1,Stuart_Review,Needs_Review,Needs_Old_Review,QMC_Review} is a projector method that evolves a trial wave function in imaginary time and projects out the ground-state wave function. For practical simulations of fermions, the fixed-node approximation is introduced, which depends only on the set of electronic positions where a trial wave function is equal to zero.  
  Wave function forms that go beyond the Born-Oppenheimer approximation are not difficult to generate, but finding accurate forms is an open question that has generated  much recent interest~\cite{cederbaum1,cederbaum12,Tubman_ECG,boent,gross2014}.  For FN-DMC the treatment of electron-ion wave functions requires minimal changes.  The main differences are seen in that one must use a different form of the trial wave function that includes the ions and additionally the  kinetic energy term for the ions must be included in the Hamiltonian.   The quality of these simulations depends on the nodal surface which  is determined by the  coordinates of both the electrons and ions simultaneously.  This is a signficantly different approximation from other methods that can simulate Hamiltonians beyond the Born-Oppenheimer approximation\cite{mitroy2013,Kurt_XCNEO-HF,Sharon_NEO-DFT,kerley2013}.

For clamped-ion simulations, the fixed-node approximation has been tested extensively with many different types of wave functions.  When the trial wave function has the same nodal surface as the exact ground-state wave function, FN-DMC yields the unbiased ground-state energy.  Approximate nodal surfaces can be generated through wave function optimization.  Approximate nodal surfaces have been tested on a wide range of systems and provide results comparable to the state of the art in \textit{ab initio} simulations.~\cite{Stuart_Review,rothstein1,grossman1,Yang2015,Tubman_Release} In addition, the energies generated with FN-DMC are variational with respect to the ground-state energy.

With the exception of some very recent research~\cite{Tubman_ECG,Yang2015}, there has been little work in treating nonadiabatic simulations of ground state wave functions with FN-DMC. Seminal work using FN-DMC for electron-ion simulations focused on condensed systems of Hydrogen~\cite{Ceperley_1987,Natoli_1993,Natoli_1995}.  One of the most well known studies of a molecular system is by Chen and Anderson   on  the H$_{2}$ molecule~\cite{chen1995}. 
  The wave function they used is specified completely in terms of relative coordinates and only a few variational parameters. Since the terms used in the wave function depend only on relative distances and are rotationally symmetric, the ions and electrons are free to rotate and translate in space. 
The success of H$_{2}$ is misleading because it can always be simulated exactly with diffusion Monte Carlo, as it has no sign problem~\cite{Tubman_Release}.  This implies that there are no systematic biases in the DMC simulation, and thus the best DMC results can be considered those that have the smallest error bar.  Therefore the variance of the local energy and the computational expense needed to evaluate the trial wave function are the important factors for generating a wave function to simulate H$_{2}$.  For most other systems, the fixed-node approximation has to be employed, and then the quality of the nodal surface becomes a crucial aspect that determines the accuracy of a simulation.  Thus the challenge of performing FN-DMC on such systems is to find good wave function forms that generate nodal surfaces of high quality.


\subsection{Electron-Ion Wave Function}
 There are several forms in which one might try to build a wave function for electron-ion systems.  The previously discussed wave functions used for H$_{2}$ are not easy to scale up to larger systems in which defects in the nodal surface can cause biases in the final results.   
In previous work we have considered several different wave function forms that make use of standard clamped-ion quantum chemistry methods~\cite{Tubman_ECG}. 
We considered three classes of wave functions that are progressively more accurate as follows:
\begin{align}
\Psi(r,R) =& e^{J(r,R)}\phi(R)\sum_{i}c^{*}_{i} D_{i}(r) \label{eqn:wfs1}\\
\Psi(r,R) =&e^{J(r,R)}\phi(R)\sum_{i}c^{*}_{i} D_{i}(r,R^{*}) \label{eqn:wfs2}\\
\Psi(r,R) =& e^{J(r,R)}\phi(R)\sum_{i}c^{}_{i}(R) D_{i}(r,R), \label{eqn:wfs3}
\end{align}
where $r$ refers to the coordinates of all the electrons and $R$ to those of all the ions.  $J(r,R)$ is the Jastrow term which involves variational parameters that correlate the quantum particles and additionally  enforce cusp conditions on the wave function.  $\phi(R)$ is the nuclear part of the wave function. The final terms correspond to determinants $D$ and the corresponding coefficients $c_i$. The $*$ denotes how these terms are evaluated, as will be discussed. 

The nuclear part of the wave function is chosen to be a simple product of gaussian functions over each nucleus pair: 
\begin{align}
\phi(R) \propto \prod_{\substack{i \\ i<j}} e^{-a_{ij}\left(|R_{i}-R_{j}|-b_{ij}\right)^2}, 
\end{align}
where $a$ and $b$ are optimizeable parameters. In our calculations $a_{ij}$ has only a single optimized value $a$, and for $b_{ij}$ we use the Born-Oppenheimer equilibrium distance between the species involved.

The terms in these wave functions involve very specific calculations that are performed and optimized in both quantum chemistry codes and quantum Monte Carlo codes.  
The determinant terms, $c_{i}^{*}D_{i}(r) $, $c_{i}^{*}D_{i}(r,R^{*}) $, and $c_{i}^{}D_{i}(r,R)$ differ based on how we optimize the determinant coefficients $c$ and how we parameterize the evaluation of the determinants based on the ion coordinates $R$.   

The  wave function in Eq.~\eqref{eqn:wfs1} is the least accurate of the three wave functions and has a fixed determinant regardless of where the ions are.  The term $c^{*}$ indicates that the determinant coefficients have been optimized at the equilibrium geometry.  
Both the ionic part of the wave function ($\phi$) and the Jastrow depend on the ion positions, which is important as the Jastrow maintains the cusps between all the quantum particles.  

The problem with this type of wave function is that the accuracy is limited by the electronic nodes, which do not depend on the ion positions.
The wave function in Eq.~\eqref{eqn:wfs2} fixes this problem.  
The $c^{*}$ indicates that the determinant part of the wave function is optimized for the equilibrium ion positions, as in the previous wave function, but the term $R^{*}$ signifies that the determinant  depends on the position of the ions through the basis set.  Basis sets in molecular calculations are generally constructed from local orbitals centered around the atoms.  In these calculations a single particle orbital is written as $\theta(r) = \sum_{ji}\gamma_{j}(r-R_{i})$, where \textit{i} is an index for an ionic center, and \textit{j} is an index for a basis set element.  
In this form, wave functions depending on the ion positions are straightforward to create and optimize,
but difficulties may arise with the possible directional dependence of the single body orbitals, such as in covalent bonds~\cite{Tubman_ECG}. 

Eq.~\eqref{eqn:wfs3} represents what we expect to be the best wave function considered here, since it has explicit dependence on the ion positions for the single particle orbitals and the determinant coefficients. Essentially this amounts to recalculating a wave function completely each time the ion positions are changed.  In practice this would significantly increase the computational cost of these simulations as well as cause many technical challenges.  The main focus of this current work is to describe a technique in which this can be done efficiently for diatomic systems. 


\section{Dragged Node Approximation}

The fixed-node approximation is generally going to result in errors in the energy that overestimate the nonadiabatic effects.   This is a result of the increased complexity of optimizing wave functions for the full electron-ion system.   If the clamped-ion energies are more accurate than the electron-ion energies, then we will overestimate the nonadiabatic energy.   
It should be noted that in some cases the energies for the full electron-ion simulations are more accurate than for the corresponding clamped-ion simulations, as seen in previous benchmark comparisons of (Be,Be$^{+}$,B,B$^{+}$,C$^{+}$)~\cite{Yang2015}.  While it does appear that in some cases the nonadiabatic simulations are as good as or more accurate than our clamped-ion simulations, this is less likely for molecular systems in which the ions can move relative to each other.    

Our recent simulations with quantum Monte Carlo have used a particular type of nodal structure which is called the  dragged-node approximation~\cite{Tubman_ECG,Yang2015}.
This approximation can be used for wave functions in the form of Eq.~\eqref{eqn:wfs2} in which  we start by generating a wave function defined at the equilibrium geometry. When the ions change position the wave function changes based on the basis set dependence of the ion coordinates. The change in the wave function causes a corresponding change in the nodes. The dragged-node approximation is completely variational when used in FN-DMC.  
For systems that do not show strong nonadiabatic behavior the dragged-node approximation should yield excellent results. 
 It was surprising that the energy contribution from nonadiabatic effects in our previous FN-DMC calculation of the CH molecule~\cite{Yang2015} was larger than expected, indicating that we might need to use better wave function forms to accurately simulate the energy for CH.   

\section{Improving wave functions}

The wave function in Eq.~\eqref{eqn:wfs3} is much more general than what we included in our previous studies but is more difficult to generate.  In general it is not feasible to do a full wave function evaluation for each new configuration of the ions.  However, for diatomic molecules it is feasible to precompute and optimize wave functions at different distances and then use the precomputed wave functions in order to interpolate wave function amplitudes at other ion positions.  There are several different ways this can be done.   The first approach we considered is to use a grid with regard to the distance between the ions and calculate a fully optimized electronic wave function at each grid point.  Then one would  evaluate the electronic wave function at each grid point and use an interpolation scheme to determine the full wave function.  This would be multiplied by a purely ionic wave function, as in Eq.~\eqref{eqn:wfs3}. Although this is technically a feasible way to generate improved wave functions,  we found that this approach was difficult to implement with regards to maintaining a smooth wave function.   

A second approach, for which we present results here, parameterizes the determinant coefficients as a function of the ion positions.  For a diatomic system, this corresponds to generating a 1D function for each determinant coefficient.  This is an improvement over the dragged-node approximation, as the coefficients of the determinants are allowed to change with ion distance, and can capture complicated ion dependences of the nodes.  In future work it also might be possible to extend this type of wave function to at least three particles, which would require fitting functions for the determinant coefficients in higher dimensions.


\section{ Wave function details}
The process for generating wave functions of the types in Eq.~(\ref{eqn:wfs2}) and Eq.~(\ref{eqn:wfs3}) requires the determination of several types of variational parameters.  For a wave function given by Eq.~(\ref{eqn:wfs2}), we use an initial guess for the wave function that is generated from complete active space self-consistent-field (CASSCF)~\cite{Chaban_MCSCF,Szabo} calculations using the quantum chemistry package GAMESS-US.~\cite{GAMESS} The optimized orbitals are then used in a configuration interaction singles and doubles (CISD) calculation to generate a series of configuration state functions (CSFs).~\cite{Pauncz_CSF} For the small systems LiH and BeH, a CASSCF calculation with a large active space is used in place of CISD. The multi-CSF expansion of the wave function can be expressed in the following form,
\begin{align}
\Psi_{\text{CISD}}(\vec{r};\vec{R}_o)=\sum\limits_{i=1}^{N_{\text{CSF}}}\alpha_i\phi_i(\vec{r};\vec{R}_o) =\sum\limits_{i=1}^{N_{\text{det} } }c_iD_i(\vec{r};\vec{R}_o), \label{eq:psi_gms}
\end{align}
where $\vec{r}$ refers to the spatial coordinates of all the electrons and $\vec{R}_o$ refers to the equilibrium positions of all the ions. $\phi_i(\vec{r})$ and $\vec{\alpha}=\{\alpha_1,\alpha_2,\dots\}$ are the CSFs and CSF coefficients, respectively, generated from CISD. Each CSF is a linear combination of determinants, so the wave function can be equivalently written as a linear combination of determinants (\ref{eq:psi_gms}). The Roos Augmented Triple Zeta ANO basis~\cite{roos} is used for the molecular systems except for the smallest system LiH, where the cc-pV5Z basis is used.

After the multi-CSF expansion is generated, we impose the electron-nucleus cusp condition on each molecular orbital~\cite{cusp} and add a Jastrow factor to the wave function to include electron correlation.~\cite{Kato} Our Jastrow factor contains electron-electron, electron-nucleus and electron-electron-nucleus terms. The full electronic wave function used in FN-DMC is,
\begin{align}
\psi_e(\vec{r};\vec{R})=e^{J(\vec{r},\vec{R},\vec{\beta})}\Psi_{\text{CISD}}(\vec{r};\vec{R}_o)\label{eq:psie},
\end{align}
We optimize the CSF and Jastrow coefficients, $\vec{\alpha}$ and $\vec{\beta}$, respectively, simultaneously with QMCPACK.~\cite{QMCPACK_Kim,QMCPACK_Esler} Optimization is performed with the ions clamped to their equilibrium positions. The equilibrium geometries for BeH and BH are chosen to be the ECG-optimized distances for comparison with the ECG  method, and the geometries for the rest of the hydrides are taken from experimental data.~\cite{CCCBDB}

For the second type of wave function, we consider a form of type Eq.~(\ref{eqn:wfs3}) as discussed in the previous section.  We specifically tested this wave function for the CH molecule by implementing the following additional steps. At the equilibrium C-H separation $R_e=2.1165$ a.u., we optimize the electronic wave function, which includes all determinant coefficients and a Jastrow. At two C-H separations near equilibrium $R_{\text{left}}=2.0$ a.u., $R_{\text{right}}=2.25$ a.u., we reoptimize only the determinant coefficients of the electronic wave function, keeping all the other wave function parameters fixed. For each determinant coefficient, we approximate its dependence on the distance between the ion positions ($R$) with an interpolation given by the following equation,
\begin{align}
c_i^*(R) = c_i(R_{\text{left}}) + 
\dfrac{c_i(R_{\text{right}}) - c_i(R_{\text{left}})}{R_{\text{right}}-R_{\text{left}}}\times(R-R_{\text{left}}). \label{eq:interpolation}
\end{align}

\begin{figure}[h]
\includegraphics[scale=0.5]{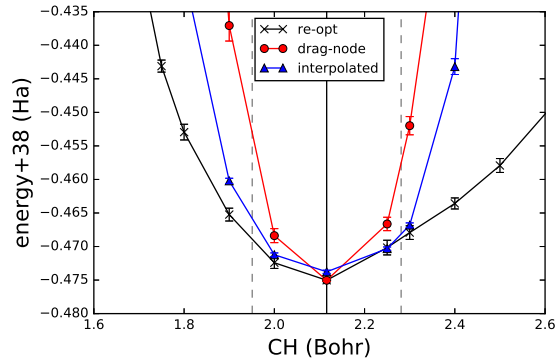}
\caption{Clamped-ion VMC total energy as a function of C-H separation using a hierarchy of wave functions. The dashed lines mark the FWHM of the distribution of C-H separation.  Within the region marked by the dashed lines it can be seen that the interpolated wave function results are a closer match to the reference 're-opt' energies than the dragged-node energies.  \label{fig:ch-cold}}
\end{figure}

We can consider a diagnostic test to determine when this type of improvement might be important.  The potential energy surface as a function of the C-H distance is plotted for several different nodal surfaces in Fig.~\ref{fig:ch-cold}. In particular we calculate clamped-ion energies that correspond to the dragged-node approximation as well as energies from a  linear interpolated wave function as given by Eq.~(\ref{eq:interpolation}). 
The reference result is obtained by re-optimizing the Jastrow factor and the determinant coefficients at every C-H separation.   The region for the most probable ion distances is indicated by the vertical dashed lines.  Over the region of important ion distances,  the potential energy surface from the interpolated wave function is improved over the dragged-node potential energy surface when compared to the fully optimized potential energy surface.  
Further away from the region of interest, both the dragged-node and the interpolated wave functions deviate signficantly from the reference data.  
This region is seldom ever sampled during our FN-DMC simulations and is not expected to affect our results. 

\section{Results and Discussion}

In our previous study, wave functions of the form in Eq.~\eqref{eqn:wfs2} were used to simulate several different molecular systems~\cite{Yang2015}. To determine the nonadiabatic contribution for each system we partition the energy into different components, which includes the clamped-ion energies, the zero point energy (ZPE) and the diagonal Born-Oppenheimer energy (DBOC).   Everything that remained we consider to be the nonadiabatic energy.   Using standard quantum chemistry tools all of the above terms can be calculated or approximated to high accuracy with the exception of the nonadiabatic energy.  As a result the nonadiabatic energy is a quantity that has not been theoretically calculated for many systems. In our previous study the nonadiabatic energy was less than 0.1~mHa for most of the systems considered.   There are two exceptions, where the nonadiabatic energy was larger, for the cases of  BH and CH molecules.   

Our new results for CH with the improved wave functions can be seen in Table~\ref{tab:energy}.  Due to the variational property of FN-DMC, it is evident that these results are improved energy over the previous best results for the CH molecule, which is not unexpected given the differences between the interpolated wave function and the dragged-node wave function as seen in Fig.~\ref{fig:ch-cold}. Our previous results showed a nonadiabatic energy of 1.9~mHa. Our new results show a nonadiabatic energy of 0.9~mHa, which can be seen for the largest determinant expansion in Table~\ref{tab:energy}.  This is consistent with our previous results, mainly that the CH molecule is somewhat nonadiabatic, even though our new estimate of the nonadiabatic energy is smaller.  For a system with a moderate amount of nonadiabatic energy, more effort is needed in generating accurate wave functions.  Improving the wave functions beyond the dragged-node approximation will lower the estimate of the nonadiabatic energy, but it is likely to remain somewhat large if the improvements of the wave function correspond to degrees of freedom beyond the Born-Oppenehimer approximation. This is what we see for CH, as the nonadiabatic energy is still  relatively large in comparison to other systems.  We note that this is still not a definitive estimate of the nonadiabatic energy, but it is likely the best estimate ever calculated for this system.

We also noticed interesting behavior that results from improving the quality of the electron nodes.  We performed clamped-ion (static) and fully nonadiabatic (dynamic) calculations using different truncation levels for the determinant expansion. The FN-DMC energy and variance for the various calculations are shown in Table~\ref{tab:energy}. As we include more determinants in our wave function, both the energy and variance of the static calculation decrease. 
However, the same does not happen for the variance of the dragged-node approximation, in which we see the surprising result that the variance increases.  This suggests that the clamped-ion wave functions are being improved to a larger extent than the dragged-node wave functions with increasing determinant number.  It is also interesting to note that for the wave functions with the smallest determinant expansion (N$_{\text{det}}$ = 35),  the variance is almost the same between the clamped-ion and dragged-node wave functions. 

The energy and variance with determinant coefficient interpolation is generally improved from our previous wave function with the dragged-node approximation.  A comparison between the dynamic runs with and without interpolation also shows that coefficient interpolation becomes more important for larger determinant expansions.  In particular, the variance improves with increasing determinant number, showing similar behavior to that of the static wave function.   

\begin{table}[h]
\begin{tabular}{llll}
\hline\hline
$N_{\text{det}}$ & Energy (Ha) & Variance (Ha$^2$) & method \\
\hline
35   & -38.4709(1) &  0.3130(5) &    static \\
35   & -38.4622(2) &  0.3169(3) &   drag \\
35   & -38.4621(2) &  0.3173(3) &  interp. \\
723  & -38.4770(1)&  0.2489(3) &    static \\
723  & -38.4667(1) &  0.334(2)~  &   drag \\
723  & -38.4679(1) &  0.2713(7) &  interp. \\
4739 & -38.4781(1) &  0.2300(4) &    static \\
4739 & -38.4676(1) &  0.334(5)~  &   drag \\
4739 & -38.4687(2) &  0.267(7)~  &  interp. \\
\hline\hline
\end{tabular}
\caption{DMC energy and variance with static ions, dynamic ions with dragged-node (``drag'') and dynamic ions with determinant coefficient interpolation (``interp.'').\label{tab:energy}}
\end{table}

\begin{figure}[h]
\includegraphics[scale=0.5]{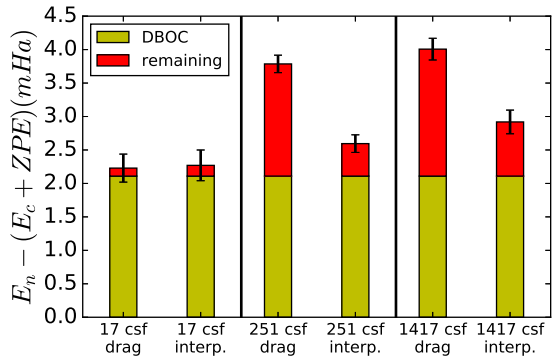}
\caption{Nonadiabatic energy of CH with and without determinant coefficient interpolation.  The wave function ``interp'' denotes that the determinant coefficients depend on C-H separation through linear interpolation. For the largest two determinant expansions a more significant contribution from nonadiabatic effects is observed than the smallest determinant expansion. \label{fig:ch-interp} }
\end{figure}

In Fig.~\ref{fig:ch-interp} we show the various contributions to the difference between the static and dynamic ground-state energies. Due to the difference in energy scales for the quantities of interest, we only plot the diagonal Born-Oppenheimer energy and the nonadiabatic energy.   To calculate the nonadiabatic energy we take the estimated zero-point energy for CH to be  6.438 mHa~\cite{Feller_Corrections}. The diagonal Born-Oppenheimer correction is estimated to be 2.11 mHa~\cite{Yang2015}. Our best result is given by the 4739 determinant interpolated wave function in Fig.~\ref{fig:ch-interp}.  Clearly, there is an increase in the nonadiabatic energy of the CH molecule that results from using the dragged-node approximation. 
 The improvement seen by using the interpolated wave function instead of the  dragged-node approximation is 1~mHa for the CH molecule; a relatively large change in the energy.  This improvement is unlikely for any of the other molecules under consideration based on our previous benchmarking.    That the dragged-node approximation produced such a large error for the CH molecule  suggests at the very least that the nodal structure of its wave function has more complex dependence on the ion configuration than the rest of the molecules under consideration.  

\begin{figure}[h]
\includegraphics[scale=0.5]{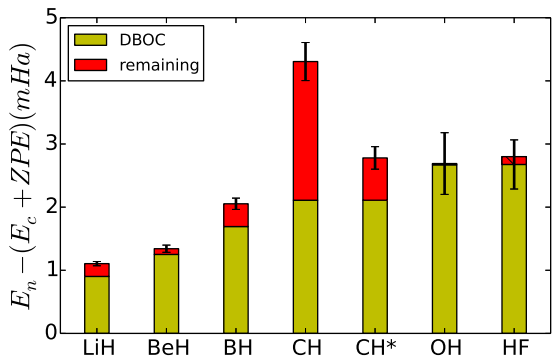}
\caption{Nonadiabatic energy of diatomic molecules.   Energies for the dragged-node calculations are taken from a previous study~\cite{Yang2015}. The best (4739 determinant) result for CH with determinant coefficient interpolation is shown with *.  Note that for all the molcules except for BH and CH the nonadiabatic energies are roughly 0.1 mHa or smaller.\label{fig:det} }
\end{figure}

Fig.~\ref{fig:ch-interp} also reveals that the nonadiabatic energy is only observed with the large determinant expansions.  There are several possible explanations for this.  It is possible we are optimzing the static wave function signficantly better than the electron-ion wave function.  There is some indication of this from the variance of the dragged-node approximation, but this is less evident for the interpolated wave function.  Another possible explanation is that only when the wave function is highly optimized do significant changes arise in the wave function amplitudes with regard to ion positions.  A related effect is that large fluctuations of the ion distance can be surpressed if the wave function and the related nodal surface is not well optimized at large ion distances.  Such effects can be mitigated altogether with the interpolated wave function approach, and are likely to be surpressed with increasing the number of determinants for the electronic part of the wave function, even for the dragged-node wave function. In Fig.~\ref{fig:det}, we compare our new results for CH with the nonadiabatic contributions from previous work. It is evident that the CH nonadiabatic energy is still much larger than all the other molecular systems. 

\section{Conclusion}

In this work, we demonstrated a new approach for generating electron-ion wave functions for diatomic systems. This approach is more accurate than those used in previous quantum Monte Carlo work.  These wave functions are generated from highly accurate clamped-ion quantum chemistry techniques, from which the derived nodes can be much more complex than those given by the dragged-node approximation.  We have specifically considered the nonadiabatic energy in the CH molecule and show that even with the improved wave function there is still a slightly larger contribution from nonadiabatic effects. Further calculations are possible to improve our results here, such as release node calculations. However, it is of interest to start testing our wave function forms on larger systems and to develop further understanding of what types of nonadiabatic effects can be captured with our current formalism.

\section{Acknowledgment}
N.T. was supported through the Scientific Discovery through Advanced Computing (SciDAC) program funded by the U.S. Department of Energy, Office of Science, Advanced Scientific Computing
Research, and Basic Energy Sciences.  This work used the Extreme Science and Engineering Discovery Environment (XSEDE), which is supported by National Science Foundation grant number ACI-1053575. D.C. and Y.Y. were supported through DOE grant DE-NA0001789. Y.Y. also acknowledges the computational science and engineering (CSE) fellowship from University of Illinois Urbana-Champaign. S.H.-S. acknowledges support by the National Science Foundation under CHE-13-61293. We used resources of the Oak Ridge Leadership Computing Facility (OLCF) at the Oak Ridge National Laboratory, which is supported by the Office of Science of the U.S. Department of Energy under Contract No. DE-AC05-00OR22725.

\bibliography{ref}
\end{document}